\documentclass[%
 reprint,
 superscriptaddress,
 bibnotes,
 amsmath,amssymb,
 aps,
 prl,
]{revtex4-2}

\usepackage{graphicx}
\usepackage{dcolumn}
\usepackage{bm}
\usepackage{color}
\usepackage{psfrag}
\usepackage{multirow}
\usepackage{booktabs}
\usepackage{rotfloat}
\usepackage{hyperref}

\renewcommand{\parallel}{\mathrel{/\mkern-5mu/}}
\makeatletter
\newcommand{\notparallel}{%
\mathrel{\mathpalette\not@parallel\relax}%
}
\newcommand{\not@parallel}[2]{%
\ooalign{\reflectbox{$\m@th#1\smallsetminus$}\cr\hfil$\m@th#1\parallel$\cr}%
}

\renewcommand{\i}{\mathrm{i}}
\renewcommand{\r}{\mathrm{r}}
\renewcommand{\t}{\mathrm{t}}
\renewcommand{\c}{\mathrm{c}}
\renewcommand{\k}{\mathbf{k}}
\renewcommand{\S}{\mathbf{S}}
\newcommand{\g}{\mathrm{g}}
\newcommand{\m}{\mathrm{m}}

\newcommand{\pati}[1]{\reversemarginpar\hspace{0pt}\marginpar{\hspace{-2.2cm}\begin{minipage}{2.8cm}\centering\vspace{\baselineskip}\textcolor{magenta}{\textit{\phantom{#1}}}\end{minipage}}}

\makeatother

\begin{document}
\preprint{APS/123-QED}

\title{Generalized Total Internal Reflection at Dynamic Interfaces}

\author{Zhiyu Li}
\email{lizhiyu@stu.xjtu.edu.cn}%
\affiliation{
 State Key Laboratory of Electrical Insulation and Power Equipment, School of Electrical Engineering, Xi’an Jiaotong University, Xi’an, Shaanxi, China
}%
\affiliation{
 Department of Electrical Engineering, KU Leuven, Leuven, Belgium
}%
 
\author{Xikui Ma}
\affiliation{
 State Key Laboratory of Electrical Insulation and Power Equipment, School of Electrical Engineering, Xi’an Jiaotong University, Xi’an, Shaanxi, China
}%
 
\author{Amir Bahrami}%
\affiliation{
  Department of Electrical Engineering, KU Leuven, Leuven, Belgium
}%

\author{Zo{\'e}-Lise Deck-L{\'e}ger}%
\affiliation{
  Department of Electrical Engineering, Polytechnique Montr{\'e}al, Montr{\'e}al, Quebec, Canada
}%

\author{Christophe Caloz}
\affiliation{
  Department of Electrical Engineering, KU Leuven, Leuven, Belgium
}%


\date{\today}
              
\begin{abstract} 
Recent research developments in the area of spacetime metamaterial structures and systems have raised new questions as to how the physics of fundamental phenomena is altered in the presence of spacetime modulation. In this context, we present a generalized and comparative description of the phenomenon of total internal reflection (TIR) at different dynamic interfaces. Such interfaces include, beyond the classical interfaces corresponding to the boundaries of moving bodies (moving interface -- moving matter systems), interfaces formed by a traveling-wave step modulation of an electromagnetic parameter (e.g., refractive index) (moving interface -- stationary matter systems) and fixed interfaces between moving-matter media (stationary interface -- moving matter systems). We first resolve the problem using the evanescence of the transmitted wave as the criterion for TIR and applying the conventional technique of relative frame hopping (between the laboratory and rest frames), which results in closed-form formulas for the relevant critical (incidence, reflection, phase refraction and power refraction) angles. We then introduce the concept of \emph{catch-up limit} between the dynamic interface and the transmitted wave as an alternative criterion for the critical angle. We use this approach both to analytically verify the critical angle formulas, further validated by full-wave (FDTD) analysis, and to explain the related physics, using Fresnel-Fizeau drag and spacetime frequency transition considerations. These results might find various applications in ultra-fast optics, gravity analogs and quantum processing.
\end{abstract}
\maketitle
\pati{Total Internal Reflection}
    \emph{Total internal reflection (TIR)} is the optical phenomenon according to which light impinging on an interface between two media is totally reflected for incidence angles beyond a threshold value, the \emph{critical angle}, which depends on the properties of the two media. This phenomenon has a long history. It was initially discussed in the XIV$\textsuperscript{th}$ century by Theodoric of Freiberg as one of the causes of the rainbow effect~\cite{freiberg1914Rainbow}. It was then revisited by several scientists from the XVII$\textsuperscript{th}$ to the XIX$\textsuperscript{th}$ century. Christiaan Huygens explained TIR in terms of his wave theory of light and identified the existence of the critical angle in his \textit{Treatise on Light} in 1690~\cite{huygens1690traite}. A few years later, Isaac Newton reinterpreted TIR from the perspective of his corpuscular theory of light and observed that the effect was frustrated (transmission occurred again) when two prisms were put in close proximity to each other in his book \textit{Opticks}~\cite{newton1704opticks}. The exact formula of the critical angle (between two isotropic media) was established only about one century later by Pierre-Simon Laplace~\cite{buchwald1989rise}. Finally, shortly afterwards, Augustin-Jean Fresnel discovered the TIR-related subtle phenomena of phase shift and polarization transformation~\cite{fresnel1817memoire}. Nowadays, TIR is an ubiquitous effect in optical technology, where it underpins a myriad of systems, such as optical waveguides, cavities, beam splitters, surface plasmonic couplers, photonic crystals, dielectric metamaterials, light-emitting diodes, scanning near-field optical microscopes, biological molecule detectors, quantum-information processing and communication systems. 
    
\pati{Statics to Dynamics}
    The vast majority of the TIR effects and devices mentioned above pertain to \emph{stationary}, or \emph{static}, systems, i.e., medium-interface configurations without any motion whatsoever. Making the medium-interface system \emph{moving}, or \emph{dynamic}, by using either moving-matter media~\footnote{The term `moving-matter medium' is used here in contrast with the term `moving-\emph{perturbation} medium', which involves the motion of stacked step modulation interfaces, as opposed to atoms and molecules~\cite{GSTEM_arXiv}.} or moving interfaces or both, naturally enriches the related physics and creates a potential for novel applications. Such systems are not limited to \emph{moving matter}, involving a moving body, with comoving atoms and molecules, that presents a moving interface to incident light (moving matter and moving interface)~\cite{yeh_reflection_1965,tsai1967wave,daly1967energy,shiozawa1968general,huang1994reflection}. They also include \emph{moving perturbation}, without net transfer of matter, in the form of a traveling-wave modulation of an electromagnetic parameter, such as for instance the refractive index (moving interface and stationary matter)~\cite{deck2018wave,caloz2019spacetime1,caloz2019spacetime2}, and \emph{moving-matter media separated by a fixed boundary} (moving matter and stationary interface)~\cite{censor1969scattering,vehmas2014transmission,taravati2017nonreciprocal,deck2021electromagnetic}. For terminological simplicity and analogical help, we refer to these systems as the `moving-truck', the `domino-chain' and the `conveyor-belt' systems, respectively.
    
\pati{Contribution}    
    This paper presents a generalized and comparative description of the phenomenon of TIR for the `moving-truck', `domino-chain' and `conveyor-belt' interfaces. First, it depicts these systems and identifies the relevant physical quantities. Second, it addresses the related problems using the relativity approach of frame hopping based on the TIR criterion of evanescent transmitted wave. Then, it introduces the alternative TIR criterion of `catch-up' between the interface and the transmitted wave, and leverages it to explain the physics of the problem. Next, it elaborates momentum diagrams to bridge the transmitted wave and the incident wave. Finally, it plots the critical angles versus velocity and explains them using the catch-up and momentum diagrams.

\begin{figure}[ht]
    \centering
    \psfrag{A}[cc][]{\textrm{(a)}}
    \psfrag{B}[cc][]{\textrm{(b)}}
    \psfrag{C}[cc][]{\textrm{(c)}}
    \psfrag{D}[cc][]{\textrm{(d)}}
    \psfrag{x}[cc][]{$x$}
    \psfrag{y}[cc][]{$y$}
    \psfrag{z}[cc][]{$z$}
    \psfrag{b}[ct][]{$z_0$}
    \psfrag{h}[ct][]{$vt$}
    \psfrag{p}[ct][]{$vt$}
    \psfrag{q}[ct][]{$v(t-\Delta t)$}
    \psfrag{n}[cc][]{$n_1$}
    \psfrag{m}[cc][]{$n_2$}
    \psfrag{M}[cc][]{$n_2(\mathbf{k})$}
    \psfrag{i}[cc][]{$\boldsymbol{\psi}_\i$}
    \psfrag{t}[cc][]{$\boldsymbol{\psi}_\t$}
    \psfrag{r}[cc][]{$\boldsymbol{\psi}_\r$}
    \psfrag{c}[cc][]{$\theta_{\i\c}$}
    \psfrag{d}[cc][]{$\theta_{\t\c}$}
    \psfrag{e}[cc][]{$\theta_{\r\c}$}
    \psfrag{f}[cc][]{$\theta_{\t\c}^{\k}$}
    \psfrag{l}[cc][]{$\theta_{\t\c}^{\S}$}
    \psfrag{k}[cc][]{$\mathbf{k}$}
    \psfrag{s}[cc][]{$\mathbf{S}$}
    \psfrag{v}[cc][]{\color[RGB]{237,28,36}$v$}
    \psfrag{V}[cc][]{\color[RGB]{237,28,36}$v_{\m}$}    
    \includegraphics[width=8.6cm]{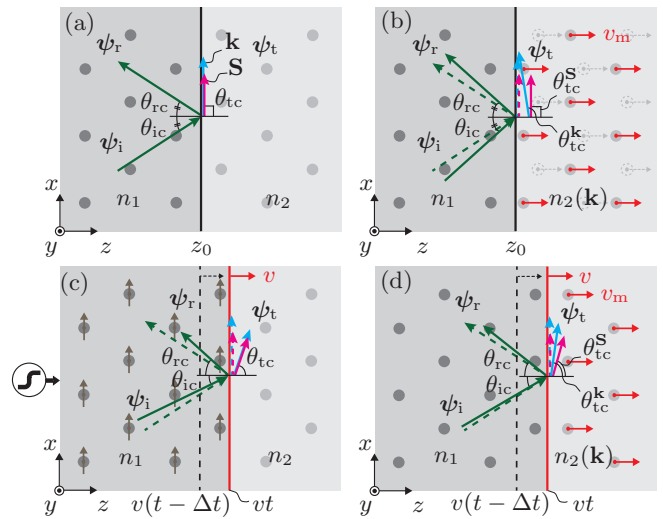}
    \caption{\label{fig:types_interf} Types of dynamic interfaces, with scattering at the critical angle. (a)~Stationary interface between stationary media (static system). (b)~Stationary interface with a moving-matter medium (conveyor-belt system). The dashed circles represent the previous positions of the moving atoms and molecules. (c)~Moving interface between stationary media (domino-chain system)~\cite{Li_MTM_09_2022}. (d)~Moving interface with a moving-matter medium (moving-truck system). It is assumed throughout the paper that $v=v_{\m}$. The green, blue and magenta dashed arrows in (b)-(d) represent the waves in medium 1, the $\mathbf{k}$ vector and $\mathbf{S}$ vector of transmitted wave in medium 2 in (a), respectively.}
    \end{figure}

\pati{Overview}
    Figure~\ref{fig:types_interf} depicts the four dynamic systems of interest and the related scattering phenomenology at the critical angle. In all the cases, the interface is sandwiched between two media, the incidence medium, labeled 1, and the transmission~\footnote{The term ``transmission'' really represents medium 2 in Fig.~\ref{fig:types_interf}, but does not necessarily imply wave transmission, as will be seen later.} medium, labeled 2. In the absence of any motion, both media are isotropic, with respective refractive indices $n_1$ and $n_2$, and $n_1>n_2$. The incidence medium is always motionless, while the interface may move at the velocity $v$ and the transmission medium may move at the velocity $v_\m$, where we will mostly consider $v_\m=v$. The subscripts `i', `r' and `t' denote the incident, reflected and transmitted waves, respectively, and the subscript `c' stands for ``critical'', and the superscripts `k' and `S' respectively refer to the wave vectors, $\mathbf{k}$, and Poynting vectors, $\mathbf{S}$. We  shall first describe the structure of these systems and later establish the theory of total reflection that applies to them.

\pati{One-by-One Description}
    Figure 1(a) shows a static system, which is constituted of a stationary interface between two stationary media. Although not dynamic, this system is shown here for reference and comparison with the following dynamic systems.  Figure~\ref{fig:types_interf}(b) represents the conveyor-belt system. In this system, the atoms and molecules composing the transmission medium (medium 2) move perpendicularly to a stationary interface, which separates the transmission medium from a stationary incidence medium (medium 1). The system is thus akin to a conveyor belt juxtaposed to a stationary object. The motion of the atoms and molecules produce a \emph{Fresnel-Fizeau drag} that makes medium 2, assumed isotropic at rest, \emph{bianisotropic} with respect to the laboratory frame (fixed $xyz$ coordinate system), $K$. As a consequence, the corresponding refractive index depends on the propagation direction, $n_2(\mathbf{k})$, and the vectors $\mathbf{k}$ and $\mathbf{S}$ are non-parallel to each other~\cite{kong1990electromagnetic,jackson1999classical}. Figure~\ref{fig:types_interf}(c) describes the domino-chain system. In this case, the interface is formed by a moving modulation step between two stationary media, i.e., two media involving no net transfer of matter. The system is akin to a chain of falling domino tiles, with its interface corresponding to the moving point between the fallen and standing titles. The domino-chain system can be both subluminal and superluminal~\cite{deck2019uniform}. However, it turns out that TIR cannot occur in the superluminal regime (Sec.~I in~\cite{supp_mat}), and we will therefore hereafter restrict our attention to the subluminal regime. Finally, Fig.~\ref{fig:types_interf}(d) shows the moving-truck system, which is the most common moving-matter system. Structurally, the moving-truck system, with both moving matter and interface, may be considered as a combination of the conveyor-belt system [Fig.~\ref{fig:types_interf}(b)], where only matter moves, and the domino-chain system [Fig.~\ref{fig:types_interf}(c)], where only the interface moves.

\pati{New angles with dynamic interfaces}
    In a static system, the critical angle, $\theta_{\i\c}$, is defined as the angle beyond which the incident wave is totally reflected at the interface, and this occurs when the transmitted wave propagates exactly along the interface, i.e., at $\theta_{\t\c}=\pi/2$, as illustrated in Fig.~\ref{fig:types_interf}(a). In contrast, in a dynamic interface system, the motion of matter or/and perturbation alters the momentum of the system and the angles must generally change to ensure the conservation of the tangential momentum. We must therefore introduce -- and find! -- the additional (nontrivial) angles $\theta_{\t\c}^{\k}$, $\theta_{\t\c}^{\S}$ and $\theta_{\r\c}$, indicated in Fig.~\ref{fig:types_interf}(b)-(d), which denote the critical transmission phase angle, the critical transmission power angle and the critical reflection (phase or power) angle, respectively. Moreover, we need to generalize the static condition $\theta_{\t\c}=\pi/2$ to the more general and more fundamental condition that the transmitted wave be \emph{evanescent} in the direction perpendicular to the interface.
    
\pati{Frame-hoping derivation of the critical angles}
    To find the critical angles $\theta_{\i\c}$, $\theta_{\r\c}$, $\theta_{\t\c}^{\k}$ and $\theta_{\t\c}^{\S}$, we shall apply the frame-hopping technique~\cite{Vanbladel_RE_1984} associated with the theory of relativity, where we define the rest frame, $K'$, as the frame of the interface, whether it is moving [Fig.~\ref{fig:types_interf}(c) and (d)] or not [Fig.~\ref{fig:types_interf}(b)]. First, we enforce the conventional (static) boundary conditions in $K'$, where the interface is stationary; this yields $\omega_{\i}'=\omega_{\r}'=\omega_{\t}'$ and $k_{x_{\i}}'=k_{x_{\r}}'=k_{x_{\t}}'$. Second, we transform these relations to the $K$ frame, using the inverse spectral Lorentz transformations $\omega'=\gamma\left(\omega \mp \beta c k_z\right)$, $ k_z'=\gamma\left(k_z \mp \beta\omega/c\right)$ and $k_x'=k_x$, where $\gamma=1/\sqrt{1-\beta^2}$ with $\beta=v/c$~\cite{kong1990electromagnetic} [NB: $\beta=0$ in Fig.~\ref{fig:types_interf}(b)], and obtain $\omega_{\i}-\beta c k_{z_{\i}}=\omega_{\r}+\beta c k_{z_{\r}}=\omega_{\t}-\beta c k_{z_{\t}}$ and $k_{x_{\i}}=k_{x_{\r}}=k_{x_{\t}}$. Third, we substitute these relations into the dispersion relations for two media, i.e., $k_{z_{\i,\r}}^2+k_{x}^{2}=\left(\omega_{\i,\r}n_1/c\right)^{2}$ for medium~1 (isotropic) and $\left(k_{z_{\t}}-\omega_{\t} \chi_2/c  \right)^{2}+\alpha_2 k_{x}^{2}=\left(\alpha_2 n_2 \omega_{\t}/c\right)^{2}$, with $\alpha_2=(1-\beta_\m^{2})/(1-\beta_\m^{2} n_2^{2})$ and $\chi_2=\beta_\m(1-n_2^{2})/(1-\beta_\m^{2} n_2^{2})$, for medium~2 [bianisotropic (Sec.~II in~\cite{supp_mat}), except when matter does not move, as in Fig.~\ref{fig:types_interf}(c), where $\beta_\m=0$]. Finally, we search for the limit where $k_{z_{\t}}$ become complex (evanescence regime) in the resulting equations and solve for the angles. The results are given in Table~\ref{tab:cri_ang} (see Sec.~IV in~\cite{supp_mat} for detailed derivations). In the stationary interface systems (first row), the transmission power angles ($\theta_{\t\c}^{\S}$) always equal $\pi/2$, whereas in the moving-interface systems (second row), they vary with the velocity of the interface ($v=\beta c$). At the same time, $\k \parallel \S$, and hence $\theta_{\t\c}^{\k}=\theta_{\t\c}^{\S}$, in the stationary matter systems (first column), whereas $\k \notparallel \S$ and $\theta_{\t\c}^{\k}\neq\theta_{\t\c}^{\S}$ in the moving-matter systems (second column). 
    \begin{table*}[ht]   \caption{\label{tab:cri_ang}Critical angle formulas for different dynamic interfaces.}
    \begin{tabular}{c|cc} 
    \hline\hline
     & stationary matter & moving matter \\
     \hline
     & \begin{minipage}[c]{0.34\textwidth}
        \smallskip \textrm{\sc Static System}
    \end{minipage} &  \begin{minipage}[c]{0.57\textwidth}
        \smallskip \textrm{\sc Conveyor-belt System}
    \end{minipage} \\
    {\begin{minipage}[c]{0.09\textwidth}\vspace{-0.8cm}\textrm{ stationary \\interface} 
    \end{minipage}}
     &
    {\begin{minipage}[c]{0.34\textwidth}
    \begin{subequations} \label{eq:crit_conventional}
        \begin{equation}
            \theta_{\i\c}=\arcsin\left(n_2/n_1\right)
        \end{equation}
        \begin{equation}
            \theta_{\r\c}=\theta_{\i\c}
        \end{equation}
        \begin{equation}
            \theta_{\t\c}^{\S}=\theta_{\t\c}^{\k}=\pi/2
        \end{equation}
        \begin{equation*}
            \phantom{\left(\frac{\chi_{2}}{\sqrt{\alpha_2 n_2^2+\chi_2^2}}\right)}
        \end{equation*}
    \end{subequations}
    \smallskip
    \end{minipage}}
    &
    {\begin{minipage}[c]{0.57\textwidth}
    \begin{subequations} \label{eq:crit_conv_belt}
        \begin{equation}
            \theta_{\i\c}=\arcsin\left(\sqrt{\alpha_2}n_2/n_1\right)
        \end{equation}
        \begin{equation}
            \theta_{\r\c}=\theta_{\i\c}
        \end{equation}
        \begin{equation}
            \theta_{\t\c}^{\S}=\pi/2
        \end{equation}        
        \begin{equation}
            \theta_{\t\c}^{\k}=\arccos\left(\frac{\chi_{2}}{\sqrt{\alpha_2 n_2^2+\chi_2^2}}\right)
        \end{equation}  
    \end{subequations}
    \smallskip
    \end{minipage}} \\
    \hline
     & \begin{minipage}[c]{0.34\textwidth}
        \smallskip \textrm{\sc Domino-Chain System}
    \end{minipage} & \begin{minipage}[c]{0.57\textwidth}
        \smallskip \textrm{\sc Moving-Truck System}
    \end{minipage} \\
    \begin{minipage}{0.09\textwidth} \vspace{-0.6cm}
        \textrm{moving\\interface}
    \end{minipage} 
    &
    \begin{minipage}{0.34\textwidth}\vspace{-0.5cm}
    \begin{subequations} \label{eq:crit_domino}
        \begin{equation}
            \theta_{\i\c}=\arcsin\left(n_2/n_1\right)-\arcsin\left(\beta n_2\right)
            \vphantom{\theta_{\i\c}=\arcsin\left(\frac{n_2}{n_1\sqrt{1-\beta^2+\beta^2 n_2^2}}\right)-\arcsin\left( \frac{\beta n_2}{\sqrt{1-\beta^2+\beta^2 n_2^2}}\right)}
        \end{equation}
        \begin{equation}
            \theta_{\r\c}=\arcsin\left(n_2/n_1\right)+\arcsin\left(\beta n_2\right)
            \vphantom{\theta_{\r\c}=\arcsin\left(\frac{n_2}{n_1\sqrt{1-\beta^2+\beta^2 n_2^2}}\right)+\arcsin\left( \frac{\beta n_2}{\sqrt{1-\beta^2+\beta^2 n_2^2}}\right)}
        \end{equation}
        \begin{equation}
            \theta_{\t\c}^{\S}=\theta_{\t\c}^{\k}=\arccos\left(\beta n_2\right)
            \vphantom{\theta_{\t\c}^{\S}=\pi/2}
        \end{equation}
        \begin{equation*}
            \phantom{\theta_{\t\c}^{\S}=\arccos\left( \frac{\beta n_2}{\sqrt{1-\beta^2+\beta^2 n_2^2}}\right)}
        \end{equation*}
    \end{subequations}
    \smallskip
    \end{minipage}
    &
    \begin{minipage}{0.57\textwidth}
    \begin{subequations} \label{eq:crit_truck}
        \begin{equation}
            \theta_{\i\c}=\arcsin\left(\frac{n_2}{n_1\sqrt{1-\beta^2+\beta^2 n_2^2}}\right)-\arcsin\left( \frac{\beta n_2}{\sqrt{1-\beta^2+\beta^2 n_2^2}}\right)
        \end{equation}
        \begin{equation}
            \theta_{\r\c}=\arcsin\left(\frac{n_2}{n_1\sqrt{1-\beta^2+\beta^2 n_2^2}}\right)+\arcsin\left( \frac{\beta n_2}{\sqrt{1-\beta^2+\beta^2 n_2^2}}\right)
        \end{equation}
        \begin{equation}
            \theta_{\t\c}^{\S}=\arccos\left( \frac{\beta n_2 }{\sqrt{1-\beta^2+\beta^2 n_2^2}}\right)
        \end{equation}        
        \begin{equation}
            \theta_{\t\c}^{\k}=\arccos\left( \frac{\beta }{\sqrt{n_2^2+\beta^2-\beta^2 n_2^2}}\right)
        \end{equation}  
    \end{subequations}
    \smallskip
    \end{minipage}  \\
    \hline
    \hline
    \end{tabular}
    \end{table*}
   
\begin{figure*}[ht]
    \centering
    \psfrag{A}[cc][]{\textrm{(a)}}
    \psfrag{B}[cc][]{\textrm{(b)}}
    \psfrag{C}[cc][]{\textrm{(c)}}
    \psfrag{D}[cc][]{\textrm{(d)}}
    \psfrag{x}[cc][]{$x$}
    \psfrag{y}[cc][]{$y$}
    \psfrag{z}[cc][]{$z$}
    \psfrag{X}[cc][]{$\overline{k}_{x\t}$}
    \psfrag{Z}[cc][]{$\overline{k}_{z\t}$}
    \psfrag{b}[ct][]{$z_0$}
    \psfrag{h}[ct][]{$vt$}
    \psfrag{p}[ct][]{$vt$}
    \psfrag{q}[ct][]{$v(t-\Delta t)$}
    \psfrag{n}[cc][]{$n_1$}
    \psfrag{m}[cc][]{$n_2$}
    \psfrag{M}[cc][]{$n_2(\k)$}
    \psfrag{i}[cc][]{$\boldsymbol{\psi}_\i$}
    \psfrag{t}[cc][]{$\boldsymbol{\psi}_\t$}
    \psfrag{c}[cc][]{$\theta_{\i\c}$}
    \psfrag{d}[cc][]{$\theta_{\t\c}$}
    \psfrag{e}[cc][]{$\theta_{\i\c}$}
    \psfrag{f}[cc][]{$\theta_{\t\c}^{\k}$}
    \psfrag{l}[cc][]{$\theta_{\t\c}^{\S}$}
    \psfrag{1}[cc][][0.8]{$\theta_{\i\c}$}
    \psfrag{7}[cc][][0.8]{$\theta_{\r\c}$}
    \psfrag{2}[cc][][0.8]{$\theta_{\t\c}^{\k}$}
    \psfrag{3}[cc][][0.8]{$\theta_{\t\c}^{\S}$}
    \psfrag{4}[cc][][0.6]{$\Delta \theta_{\t\c}^{\S}$}
    \psfrag{5}[cc][]{$n_1$}
    \psfrag{6}[cc][]{$n_2(\mathbf{k})$}
    \psfrag{k}[cc][]{$\k$}
    \psfrag{s}[cc][]{$\S$}
    \psfrag{v}[cc][]{\color[RGB]{237,28,36}$v$}
    \psfrag{V}[cc][]{\color[RGB]{237,28,36}$v_{\m}$}
    \psfrag{E}[cc][]{\begin{minipage}{2cm}\centering{\textrm{TIR\\(evanescence)}}\end{minipage}}
    \psfrag{P}[cc][]{\text{transmission}}
    \psfrag{0}[cc][0.8]{\begin{minipage}{6cm}\centering{\textrm{normalized isofrequency ellipses\\in medium 2}}\end{minipage}}
    \psfrag{G}[cc][]{$\mathbf{v}_{\g_{\t}}$}
    \psfrag{g}[cc][]{$v_{\g z_{\t}}$
    }
    \includegraphics[width=17.8cm]{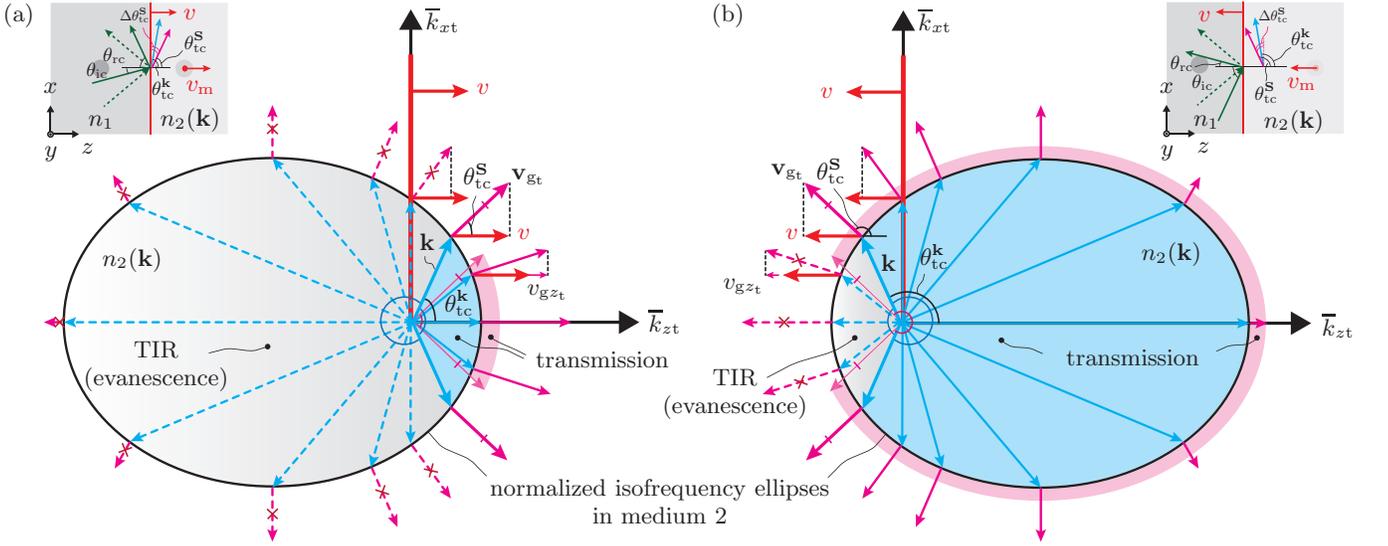}
    \caption{\label{fig:catch-up_truck} Catch-up limit between the interface and the transmitted wave, shown here for the case of the moving-truck system. (a)~Forward interface ($v=v_{\m}>0$). (b)~Backward interface ($v=v_{\m}<0$). The blue, pink and gray regions correspond to the the sectors of phase transmission, power transmission and TIR (evanescence), respectively.}
    \end{figure*}

\pati{Motivation and essence of the Catch-up condition}
    Table~\ref{tab:cri_ang} provides the formulas for the critical angles, but it does not provide insight into the physics of the TIR phenomenon in the considered systems. Therefore, we introduce here an alternative perspective of TIR, namely the \emph{catch-up limit} between the interface and the \emph{transmitted} wave -- assuming that the incident wave is always fast enough to catch up with the interface for scattering to occur. This perspective is illustrated in Fig.~\ref{fig:catch-up_truck} for the most general case of the moving-truck system, where the isofrequency curves are shifted ellipses, due to matter motion (drag). It is based on the recognition that the limit of transmission evanescence corresponds to the limit where the velocity of the interface ($v$) equals the velocity of the transmitted wave power in the direction of motion, $v_{\g z_{\t}}$, i.e., $v=v_{\g z_{\t}}$. In the forward case ($v=v_\m>0$), represented in Fig.~\ref{fig:catch-up_truck}(a), the interface moves below this limit ($v<v_{\g z_{\t}}$), so that the transmitted wave propagates in medium~2 (refraction), whereas the interface ``catches it up'', and hence makes it evanescent in medium~2, above this limit ($v>v_{\g z_{\t}}$), which corresponds to the TIR regime; the situation is inverted in the backward case, represented in Fig.~\ref{fig:catch-up_truck}(b). This procedure leads to $\theta_{\t\c}^{\S}=\arccos(v/|v_{\g_{\t}}|)$, from which we can compute $\theta_{\t\c}^{\k}$, and then $\theta_{\i\c}$ and $\theta_{\r\c}$ (Sec.~IV in~\cite{supp_mat}). 
    
\pati{Particular cases: domino-chain system and conveyor-belt system}
    The domino-chain, conveyor-belt and static systems may be further understood in terms of the the catch-up limit perspective as particular cases of the moving-truck system (see Sec.~VI in~\cite{supp_mat} for specific graphs), with centered-circle isofrequency curves in the stationary-matter systems (domino-chain and static systems) and $\theta_{\t\c}^{\S}=\pi/2$ resulting from the zero velocity of the interface in the stationary-interface systems (conveyor-belt and static systems). The problem can naturally also be solved in the even more general regime $v\neq v_{\m}$ (Sec.~V in~\cite{supp_mat}), from which the critical angle formulas in Table~\ref{tab:cri_ang} of the four systems are obtained by setting $\beta=\beta_{\m}=0$, $\beta=0$ (and $\beta_{\m}\neq 0$), $\beta_{\m}=0$ (and $\beta\neq 0$) and $\beta=\beta_{\m}\neq 0$, respectively, in the generalized formulas. 
    
    \pati{Influence of the catch-up limit on $\theta_{\i\c}$ and $\theta_{\r\c}$ in medium 1} 
    An interesting observation may be made about the incidence and reflection angles in the moving-interface systems. Applying the identity $\arccos(x)+\arcsin(x)=\pi/2$ in Table~\ref{tab:cri_ang}, we find that $\theta_{(\i,\r)\c}=\arcsin\left(n_2/n_1\right)\mp\Delta\theta_{\t\c}^{\S}$ and $\theta_{(\i,\r)\c}=\arcsin\left(n_2/(n_1\sqrt{1-\beta^2+\beta^2 n_2^2})\right)\mp\Delta\theta_{\t\c}^{\S}$, for the domino-chain and moving-truck systems, respectively, where $\Delta\theta_{\t\c}^{\S}=|\theta_{\t\c}^{\S}-\pi/2|$ is the excess power angle due to the motion of the interface. These relations reveal that the incidence and reflection critical angles are simply offset by the catch-up (power) angle. Moreover, defining a new coordinate system rotated by the angle $\Delta\theta_{\t\c}^{\S}$ towards the direction of $\mathbf{v}$ results into relations that are formally identical to the static ones for the domino-chain case and, specifically, lead hence to the concept of an \emph{effective interface} for that system (see Fig.~S1 in~\cite{supp_mat}).
    
\begin{figure}[ht]
    \centering
    \psfrag{A}[cc][]{\textrm{(a)}}
    \psfrag{B}[cc][]{\textrm{(b)}}
    \psfrag{C}[cc][]{\textrm{(c)}}
    \psfrag{D}[cc][]{\textrm{(d)}}
    \psfrag{X}[lc][][0.8]{$\overline{k}_{x}$}
    \psfrag{Z}[lc][][0.8]{$\overline{k}_{z}$}
    \psfrag{n}[cc][][0.8]{$n_1$}
    \psfrag{m}[cc][][0.8]{$n_2$}
    \psfrag{M}[cc][][0.8]{$n_2(\mathbf{k})$}
    \psfrag{N}[cc][][0.8]{\begin{minipage}{1.5cm}\centering\vspace{\baselineskip}{$n_{2\mathrm{eff}}(\mathbf{k})$\\$=n_2(\mathbf{k})$}\end{minipage}}
    \psfrag{c}[cc][][0.8]{$\theta_{\i\c}$}
    \psfrag{d}[cc][][0.8]{$\theta_{\t\c}$}
    \psfrag{e}[cc][][0.8]{$\theta_{\i\c}$}
    \psfrag{f}[cc][][0.8]{$\theta_{\t\c}^{\k}$}
    \psfrag{l}[cc][][0.8]{$\theta_{\t\c}^{\S}$}
    \psfrag{v}[cc][][0.8]{\color[RGB]{237,28,36}$v$}
    \psfrag{V}[cc][][0.8]{\color[RGB]{237,28,36}$v_{\m}$}
    \psfrag{G}[cc][][0.8]{$\mathbf{v}_{\g}$}
    \psfrag{g}[cc][][0.8]{$v_{\g z}$}
    \psfrag{w}[cc][][0.8]{\begin{minipage}{1.5cm}\centering\vspace{\baselineskip}{$n_{2\mathrm{eff}}(\mathbf{k})$\\ $\Delta\omega\neq0$}\end{minipage}}
    \includegraphics[width=8.6cm]{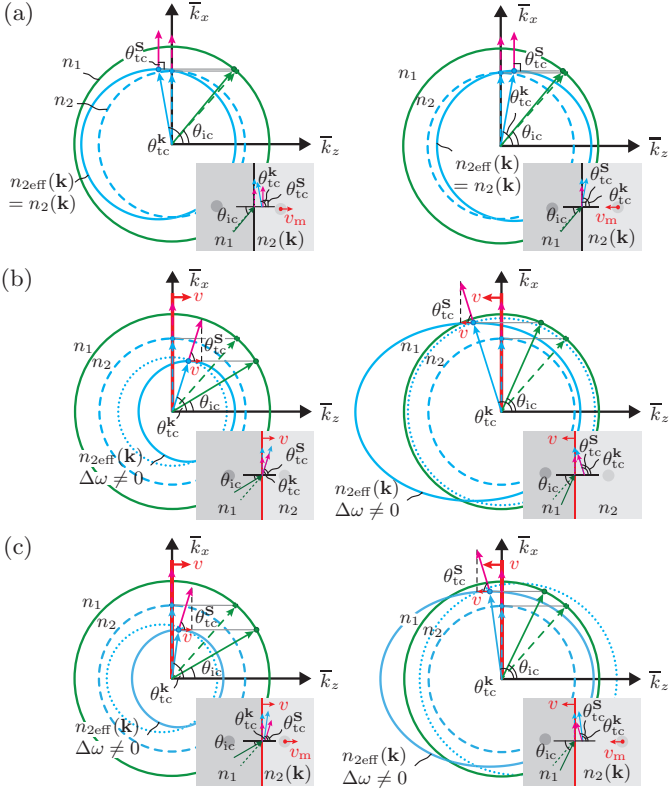}
    \caption{\label{fig:mom_surf} Momentum diagrams for the incident and transmitted waves in terms of the normalized wavenumber $\overline{k}_{\i,\t}=k_{\i,\t}/(\omega_\i/c)$, along with critical angles, for the dynamic interfaces in Fig.~\ref{fig:types_interf}, for $(n_1,n_2)=(2,1.5)$ and both positive $z$-motion (left), with $\beta=0.2$, and negative $z$-motion (right), with $\beta=-0.2$. (a)~Conveyor-belt system. (b)~Domino-chain system. (c)~Moving-truck system. In all the cases, the dashed curves correspond to the static case. In the cases (b) and (c), the dotted curves are isofrequency curves at $\theta_{\t\c}^\k$.}
    \end{figure}

    \pati{Introduction to Momentum Diagrams}
    The catch-up graphs of Fig.~\ref{fig:catch-up_truck} were instrumental for illuminating the physics of the dynamic TIR phenomenology, but they are restricted to the representation of the transmitted wave. In order to bridge the transmitted wave to the \emph{incident} wave, which is the wave associated with the most important critical angle (the source angle), Fig.~\ref{fig:mom_surf} introduces incidence-transmission \emph{momentum diagrams} (see Sec.~VII in~\cite{supp_mat} for diagrams including also reflected waves). These diagrams are dynamic generalizations of isofrequency diagrams that follow possible anisotropic frequency transformations. They also provide here a comparison means with the static case as well as a global perspective of scattering beyond the critical angles.
    
    \pati{Construction of the Momentum Diagrams}
    These momentum diagrams in Fig.~\ref{fig:mom_surf} are obtained as follows. First, we rewrite the phase-matching condition $\omega_{\i} n_1 \sin\theta_{\i}/c = \omega_{\t} n_2(\k_{\t})\sin\theta_{\t}^{\k}/c$ as $n_1\sin\theta_{\i}=n_{2,\mathrm{eff}}(\k_{\t})\sin\theta_{\t}^{\k}$, where $n_{2,\mathrm{eff}}(\k_{\t})=n_2(\k_{\t}) \omega_{\t}(\k_{\t})/\omega_{\i}$ is the effective (anisotropic and anisofrequency) refractive index of medium~2, which is constituted of the moving-interface frequency-transformation term $\omega_\t(\k_{\t})=\omega_\i[(1-\beta n_1 \cos\theta_{\i}]/[1-\beta n_2(\k_{\t}) \cos\theta_{\t}^{\k}]$ and the moving-matter bianisotropy term, $n_2(\k_{\t})$  (Sec.~VII in~\cite{supp_mat})~\footnote{Substituting the relation $\cos\theta_{\i}=\{ \beta n_2(\k_{\t})^2 {\sin^{2}\theta_{\t}^{\k}} + [1 - \beta n_2(\k_{\t}) \cos\theta_{\t}^{\k}]\sqrt{M} \}/[n_1(\beta^2 n_2(\k_{\t})^2-2 \beta n_2(\k_{\t}) \cos\theta_{\t}^{\k} +1) ]$ with $M=n_1^2[\beta^2 n_2(\k_{\t})^2-2 \beta n_2(\k_{\t}) \cos\theta_{\t}^{\k}+1]-n_2(\k_{\t})^2 {\sin^{2}\theta_{\t}^{\k}}$~\cite{GSTEM_arXiv} into the formula in the main text for $\omega_\t(\k_{\t})$, and further substituting the result into $n_{2,\mathrm{eff}}(\k_{\t})$, provides an explicit formula for $n_{2,\mathrm{eff}}=n_{2,\mathrm{eff}}(\theta_{\t}^\k)$.} Then, we plot the $n_1$ and $n_{2,\mathrm{eff}}(\k_{\t})$ curves in the $\overline{k}_{z}\overline{k}_{x}$ plane.
    
    \pati{Interface Domination in Scattering}
    It appears in Fig.~\ref{fig:mom_surf} that the $\overline{k}_z\overline{k}_x$ curves for the domino-chain system [Fig.~\ref{fig:mom_surf}(b)] and moving-truck system [Fig.~\ref{fig:mom_surf}(c)] are very similar to each other, while the curves for the conveyor-belt system [Fig.~\ref{fig:mom_surf}(a)] are very different from them. Specifically, the transmission curves, which include information on \emph{all} the scattered angles (not just the critical ones), are shifted in \emph{opposite} directions  for the former (domino-truck) and latter (conveyor) systems. The comparison of the conveyor-belt and domino-chain systems indicates that the effect of interface motion is opposite to the effect of matter motion on the transmission momentum curves; this is explained by the fact that the Fresnel-Fizeau drags in the $K'$ frame are opposite for the two systems. The fact that the moving-truck system behaves similarly to the domino-chain system, while it includes both interface and matter motion, further reveals that the motion of the interface dominates the motion of matter in the overall scattering phenomenology, apparently because scattering is more influenced by the discontinuity formed by the interface than by the media surrounding it. Thus, interestingly, a matter interface, which involves cumbersome moving parts, can essentially be replaced by a modulation interface, which would be more practical for electromagnetic devices.
    
    \pati{Equivalent TR Condition of $k_{z\t}$ becomes complex: $k_{x\t}$ becomes maximum}
    The critical angles are graphically found from the momentum diagrams upon noticing that $k_{x_{\i\c}}=\mathrm{max}(k_{x_{\t}}=\sqrt{k_{\t}^2-k_{z_{\t}}^2})$, or $\overline{k}_{x_{\i\c}}=\mathrm{max}\left(\overline{k}_{x_{\t}}=\sqrt{\overline{k}_{\t}^2-\overline{k}_{z_{\t}}^2}\right)$
    in normalized terms, is an alternative criterion to the transmission evanescence criterion. Indeed, the $k_{x_{\i}}=k_{x_{\t}}$ matching condition cannot be satisfied any more when $k_{x_{\i}}$ becomes larger than $\mathrm{max}(k_{x_{\t}})$\footnote{The incidence circle is larger than the transmission curve (not an ellipse) because $n_1>n_2$.}, which corresponds to the top of the transmission curve; this means that the transmitted wave becomes evanescent above this limit, and that this limit is therefore equivalent to evanescence TIR criterion. From this point, $\theta_{\i\c}$ is found by simply tracing a horizontal line from the top of the transmission curve to the incidence curve, and reading out the corresponding angle from the $\overline{k}_z$ axis. It appears (see Fig.~\ref{fig:mom_surf}) that, as a direct consequence of the opposite shift directions of the momentum curves explained above, the incidence critical angle is slightly increased by motion in the conveyor-belt system for both the forward and backward cases, and strongly decreased in the forward case and increased in the backward case by motion for the domino and truck systems.
    
    \pati{Extra comments on the Momentum Diagrams}   
    Note that the transmission group velocity is no longer the gradient of the effective refractive index curves in moving-interface systems [Fig.~\ref{fig:mom_surf}(b) and (c)], since these curves do not conserve frequency ($\Delta\omega\neq0$); to find this group velocity, one must plot the isofrequency curve \emph{locally}, at each $\theta_{\t}$ point of interest, as done in the figure (in dotted line), at the critical angle point, where the isofrequency curves are circles in the domino-chain system and off-centered ellipses in the moving-truck system, respectively.
    
\begin{figure}[ht]
    \centering
    \psfrag{A}[cc][]{\textrm{(a)}}
    \psfrag{B}[cc][]{\textrm{(b)}}
    \psfrag{C}[cc][]{\textrm{(c)}}
    \psfrag{D}[cc][]{\textrm{(d)}}
    \psfrag{i}[cc][]{$\theta_{\i\c}$}
    \psfrag{r}[cc][]{$\theta_{\r\c}$}
    \psfrag{s}[cc][]{$\theta_{\t\c}^{\S}$}
    \psfrag{k}[cc][]{$\theta_{\t\c}^{\k}$}
    \psfrag{I}[rc][][0.8]{$\sin^{\text{-}1}(n_2/n_1)$}
    \psfrag{R}[rc][][0.8]{$\sin^{\text{-}1}(n_2/n_1)$}
    \psfrag{S}[rc][][0.8]{$\pi/2$}
    \psfrag{K}[rc][][0.8]{$\pi/2$}
    \psfrag{p}[lc][][0.8]{$\pi/5$}
    \psfrag{b}[cc][]{$\beta$}
    \psfrag{n}[ct][]{$\frac{1}{n_1}$}
    \psfrag{m}[ct][]{$\text{-}\frac{1}{n_1}$}
    \psfrag{0}[ct][]{0}
    \psfrag{1}[cc][][0.7]{\begin{minipage}{1.5cm}\centering\vspace{\baselineskip}{\textrm{conveyor\\ belt}}\end{minipage}}
    \psfrag{2}[cc][][0.7]{\begin{minipage}{1.5cm}\centering\vspace{\baselineskip}{\textrm{domino\\chain}}\end{minipage}}
    \psfrag{3}[cc][][0.7]{\begin{minipage}{1.5cm}\centering\vspace{\baselineskip}{\textrm{moving\\truck}}\end{minipage}}
    \psfrag{4}[rc][][0.7]{\textrm{static}}
    \psfrag{H}[lc][][0.6]{\textrm{Theory}}
    \psfrag{h}[lc][][0.6]{\textrm{FDTD}}
    \includegraphics[width=8.6 cm]{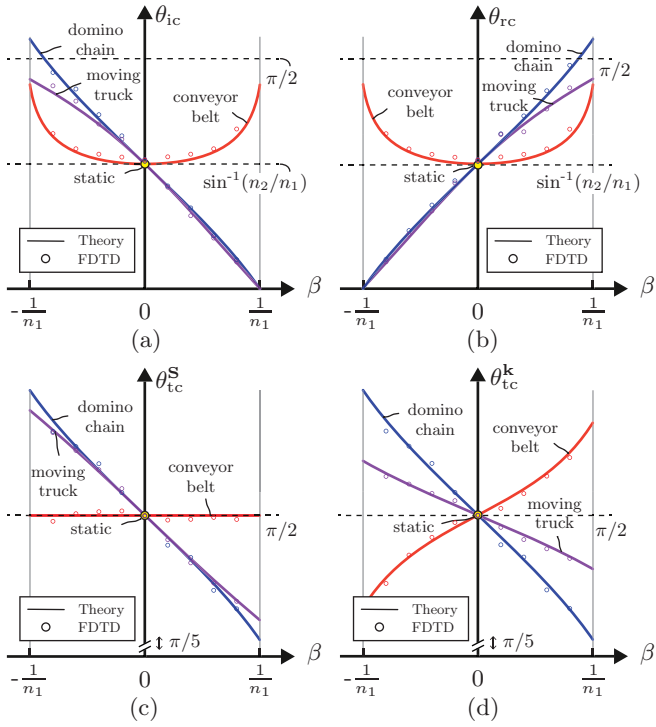}
    \caption{\label{fig:theta_beta} Critical angles versus velocity ($\beta=v/c$) for the four systems in Fig.~\ref{fig:types_interf} with the same refractive indices in Fig.~\ref{fig:mom_surf} (The curves in (c) and (d) have been downshifted by $\pi/5$ rad for visualization convenience). (a)~Incidence. (b)~Reflection. (c)~Power transmission. (d)~Phase transmission.}
    \end{figure}

\pati{Critical angles versus $\beta$: overview}
    On the basis of the physics established in conjunction with Figs.~\ref{fig:catch-up_truck} and~\ref{fig:mom_surf}, we can now investigate the dependence of the critical angles given in Table~\ref{tab:cri_ang} on the velocity of the interface or/and matter. The corresponding curves are plotted in Fig.~\ref{fig:theta_beta}~\footnote{The critical angles for the conveyor-belt system (not shown) have opposite slopes when $n_2<1$ (plasma-type medium). Moreover, the critical angles for all the dynamic systems (also not shown) remain unchanged when $n_2<0$ with $|n_2|$ unchanged (negative-index medium).}. These results are validated by Finite-Difference Time-Domain (FDTD)~\cite{taflove2005computational} simulations that combine the frame-hopping technique \cite{harfoush1990numerical,Bahrami_MTM_09_2022} and the spatial Fourier transform of the scattered fields~\cite{achouri2018space,Li_MTM_09_2022} (data given in Sec.~VIII in~\cite{supp_mat} and full-wave animations available at \cite{supp_mat_animation}).

\pati{Global Description in Fig.~\ref{fig:theta_beta}}
    Several observations and explanations are in order in Fig.~\ref{fig:theta_beta}. A global observation is that the domino-chain and moving-truck systems exhibit the same, monotonic slope responses for all the critical angles, while the responses of the conveyor-belt system are very different. This corresponds to the particular case (critical angles) of the general effects (all scattering angles) explained in conjunction with Fig.~\ref{fig:mom_surf} (opposite Fresnel-Fizeau drags and predominance of interface effect). 

    \pati{Specific Descriptions in Fig.~\ref{fig:theta_beta}}
    Specific observations in Fig.~\ref{fig:theta_beta} include the following. Figure~\ref{fig:theta_beta}(a) shows that the critical incidence angle response for the conveyor-belt system is symmetric with respect to $\beta=0$ with $\partial \theta_{\i\c}/\partial |\beta|>0$, and decreases monotonically ($\partial \theta_{\i\c}/\partial \beta<0$) in the moving-interface (domino and tuck) systems down to zero, where the incident wave cannot catch up with the moving interface any more. These incidence critical angle responses have been explained from the perspective of momentum variations in conjunction with Fig.~\ref{fig:mom_surf}. Figure~\ref{fig:theta_beta}(b) shows that the critical reflection angle responses are always the $\beta$-symmetric counterparts of the incidence responses in Fig.~\ref{fig:theta_beta}(a); this is explained by the time-reversal condition $\theta_{\r}(\beta,n_1)=\theta_{\i}(-\beta,n_1)$ in the TIR regime (only 2 propagating waves). Figure~\ref{fig:theta_beta}(c) shows the $\theta_{\t\c}^\S$ responses, whose trends behavior can be understood from the catch-up explanations in  Fig.~\ref{fig:catch-up_truck} (and Fig.~S1-S3 in \cite{supp_mat}). Finally, Fig.~\ref{fig:theta_beta}(d) shows the transmission phase angle response. Compared to the curves in Fig.~\ref{fig:theta_beta}(c), the curves for the conveyor-belt and moving-truck systems experience an anticlockwise rotation about static-system point; indeed, the Fresnel-Fizeau drag associated with matter motion in these two systems rotates the group velocity vector towards the direction of motion while not affecting the direction of wave vector, as shown in Fig.~\ref{fig:catch-up_truck}, so that one must have $\theta_{\t\c}^\k\gtrless\theta_{\t\c}^\S$ for $\beta\gtrless 0$.

    Zhiyu Li would like to acknowledge the financial support from the China Scholarship Council (CSC, No. 202106280230).

\bibliography{Reference}


\end{document}